\documentclass[pdflatex,sn-mathphys-num]{sn-jnl}% Math and Physical Sciences Numbered Reference Style
\usepackage{xspace}
\newcommand{\ptbi}{t-PtBi\(_2\)\xspace}

\usepackage{graphicx}%
\usepackage{multirow}%
\usepackage{amsmath,amssymb,amsfonts}%
\usepackage{amsthm}%
\usepackage{mathrsfs}%
\usepackage[title]{appendix}%
\usepackage{xcolor}%
\usepackage{textcomp}%
\usepackage{manyfoot}%
\usepackage{booktabs}%
\usepackage{algorithm}%
\usepackage{algorithmicx}%
\usepackage{algpseudocode}%
\usepackage{listings}%
\theoremstyle{thmstyleone}%
\theoremstyle{thmstyletwo}%

\theoremstyle{thmstylethree}%

\begin{document}

\author[1]{Andrii Kuibarov}
\author[1]{Susmita Changdar}
\author[1]{Riccardo Vocaturo}
\author[1,2]{Oleksandr Suvorov}
\author[1,3]{Alexander Fedorov}
\author[1,3]{Rui Lou}
\author[3]{Maxim Krivenkov}
\author[1,4]{Luminita Harnagea}
\author[1]{Sabine Wurmehl}
\author[1]{Jeroen van den Brink}
\author[1]{Bernd B\"uchner}
\author[1]{Sergey Borisenko}

\affil[1]{Leibniz Institute for Solid State and Materials Research, IFW Dresden, Helmholtzstrasse 20, 01069 Dresden, Germany}
\affil[2]{Kyiv Academic University, 36 Vernadsky blvd, 03142 Kyiv, Ukraine}
\affil[3]{Helmholtz-Zentrum Berlin f\"ur Materialien und Energie, BESSY II, 12489 Berlin, Germany}
\affil[4]{I-HUB Quantum Technology Foundation, Indian Institute of Science Education and Research, Pune 411008, India}

\title{Three prerequisites for high-temperature
superconductivity in \ptbi }


\abstract{Although the generic mechanism behind high-temperature superconductivity remains notoriously elusive, a set of favorable conditions for its occurrence in a given material has emerged: (i) the electronic structure should have a very high density of states near the Fermi level; (ii) electrons need to be susceptible to a sizable interaction with another degree of freedom to ensure pairing themselves; (iii) the ability to fine-tune some of the system properties significantly helps maximising
the critical temperature. Here, by means of high-resolution ARPES, we show that all three criteria are remarkably fulfilled in trigonal platinum bismuthide (\ptbi). Specifically, this happens on its surface, which hosts topological surface states known as Fermi arcs. Our findings pave the way for the stabilisation and optimisation of 
high-temperature superconductivity in this 
topological material.}

% \keywords{keyword1, Keyword2, Keyword3, Keyword4}



\maketitle

\section{Introduction}\label{sec1}

High-temperature superconductivity (HTS) continues to be a focal point for research in condensed matter physics, primarily due to its potential for technological innovations and the fundamental quesions it poses about quantum states of matter. The lack of a definitive mechanism has significantly hindered the prediction of new superconducting materials, but the remarkable properties of known superconductors are being widely exploited in various applications \cite{Shimoyama2023}, underscoring the importance of identifying common conditions that facilitate superconductivity across diverse systems.
In this context, several %critical 
prerequisites have garnered consensus among researchers as beneficial for HTS.
%fundamental for high-temperature superconductivity. 

% rv
First, a high density of states (DOS) near the Fermi level naturally increases the likelihood of superconductivity to appear even in the absence of strong coupling. This is because a significant energy gain would follow as the superconducting gap opens. An interesting example of such behavior is when van Hove singularities (VHS) are found in the close proximity of the Fermi level. These singularities correspond to critical points in reciprocal space where the DOS formally diverges to infinity. They have been, thus, associated with strong instabilities and enhanced correlations.
%The presence of Van Hove singularities significantly enhances the density of electronic states near the Fermi level, which increases the likelihood of superconductivity due to energy gain and amplification of the electronic interactions. 
In many superconductors, such as cuprates \cite{Bok2012}, 
iron-based family \cite{Borisenko2013NM}, strontioum ruthenate \cite{Yokoya1996,Noad2023}, PdTe$_2$ \cite{Kim2018}, kagome materials \cite{Luo2023}, hydrides \cite{VillaCorts2022} and twisted graphene layers \cite{Li2009,Xu2021},
these features 
%arise from the underlying crystal structure and electronic interactions and 
have demonstrated their %universality and 
significance.

Second, for electrons to form Cooper pairs, they must interact with another degree of freedom, be it lattice vibrations, spin fluctuations, or other collective states \cite{Valla1999,Bostwick2010,Lanzara2001, Dahm2009}. The nature of these interactions dictates the strength and robustness of the superconducting state. ARPES enables to directly probe these couplings \cite{Kordyuk2005}, revealing insights into self-energy effects and the energy landscape around the Fermi surface, thus uncovering the role of various interactions in favoring superconductivity.

Finally, the ability to finely tune a material’s properties is essential for optimizing critical temperature ($T_c$). Various techniques: doping the charge carriers, applying external pressure or strain, and modifying the chemical environment via intercalation or substitution, allow one to explore phase diagrams that map the interactions between different parameters. By altering these conditions, one can often enhance $T_c$ significantly, as seen in high-transition temperature systems like the iron pnictides and cuprates.

In the case of the \ptbi, where the presence of topological Fermi arcs on its surface introduces an additional layer of complexity, extensive ARPES and STM studies 
demonstrated the presence of superconducting phase \cite{Kuibarov2024,Schimmel2024,changdar_iwave, JuliaTemperatureSTM,XiaochunSTM,Hoffmann2024}. In particular (in a previous work), 
we have demonstrated that \ptbi exhibits evidence of surface-only superconductivity, with a critical temperature of approximately 15~K and a superconducting gap $\Delta$ ranging from 1.5 to 2~meV~\cite{Kuibarov2024}. This gap opens exclusively on the surface topological Fermi arcs, while the bulk remains metallic down to approximately 1~K~\cite{Veyrat2023, ShipunovPRM}. Moreover, detected structure of the order parameter clearly implies that this superconductivity is not conventional \cite{changdar_iwave}. On the other hand, STM measurements have reported regions in the sample with much larger superconducting gaps, reaching up to 20~meV~\cite{Schimmel2024, XiaochunSTM}, and critical temperatures as high as 45~K~\cite{JuliaTemperatureSTM}.
These and other open questions motivated us to initiate the present study focusing on fine details of the Fermi arcs' electronic structure on both terminations. In analogy with previously discovered families of unconventional superconductors it is also important to identify the crucial components of the electronic structure responsible for the unusual superconducting state and understand whether these components are tunable in terms of energy and momentum. 
While we found the Fermi arcs to exhibit singular behaviors on both terminations, features we believe to be crucial for superconductivity, we also observed pronounced differences between them. In particular, one termination shows good agreement between ARPES and DFT-computed energy dispersion, whereas the other displays clear signs of strong renormalization, suggesting strong coupling with other excitations.
%
%In spite of very strong localization both in momentum and energy we have found similar and diverse features of the electronic structures of the Fermi arcs on opposite terminations. While both exhibit singularities that are crucial for superconductivity, the renormalization of the underlying electronic dispersion differs drastically, implying strong variations in the coupling strength with other excitations. 
We also detect size variation of the arcs even within the same termination, which suggests the possibility to tune the system and potentially control it.
We anticipate that our findings in \ptbi will not only contribute to theoretical models but also foster experimental developments in stabilizing and optimizing topological superconductivity, opening new avenues in the quest for practical applications in quantum computing and advanced electronics.

\section{Fermi arcs on two terminations}

Trigonal PtBi$_2$ is a noncentrosymmetric Weyl semimetal, with Weyl points lying above the Fermi level, as confirmed by pump–probe ARPES measurements \cite{PtBi2_pumpProbe}. The lack of inversion symmetry leads to two distinct cleavage planes. Following Ref.~\cite{Vocaturo2024}, we will refer them to as decorated-honeycomb (DH) or Kagome-type (KT) surface termination.

\begin{figure}[H]%
\centering
\includegraphics[width=1.0\textwidth]{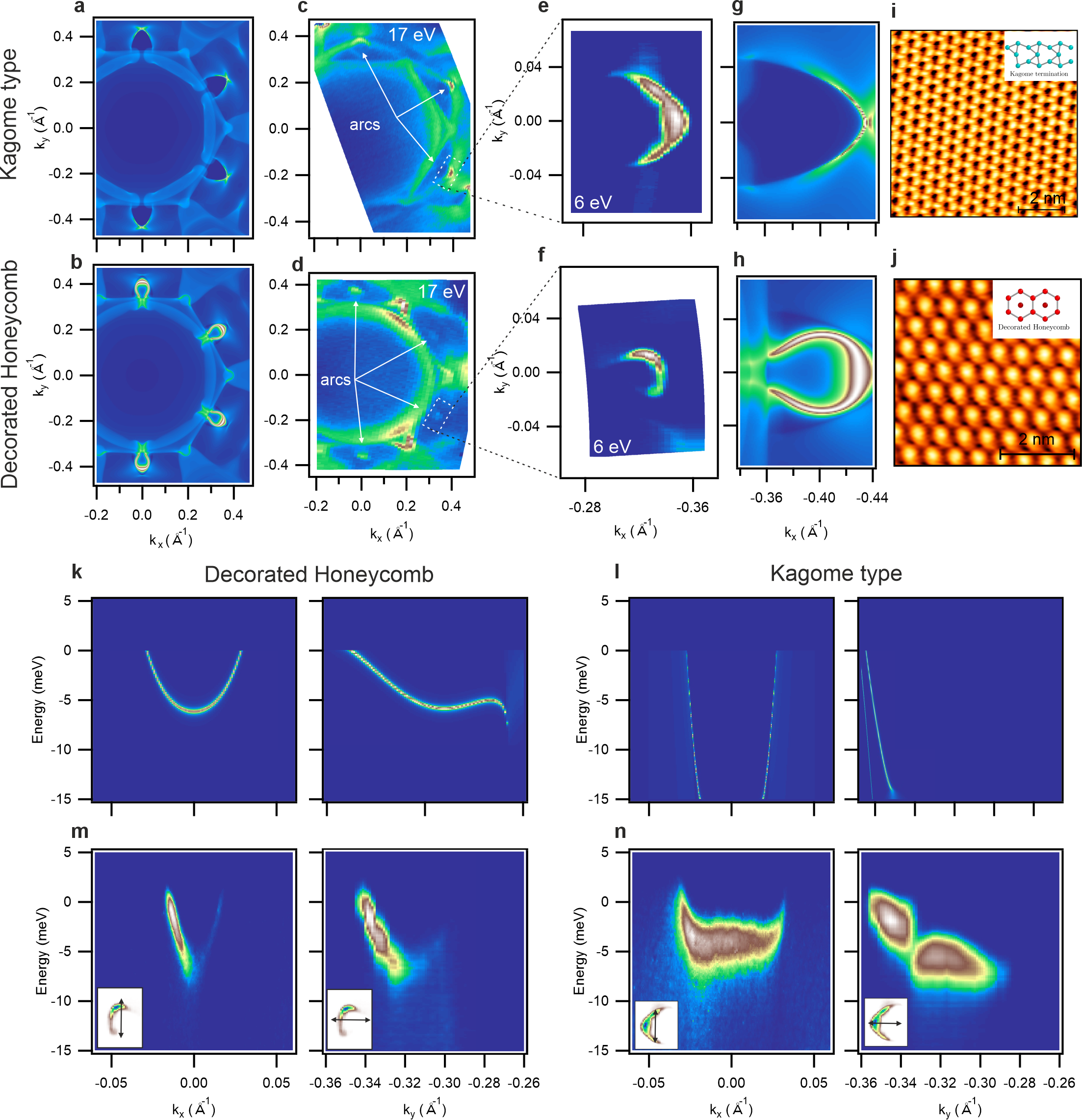}
\caption{Fermi surface maps and STM topography of two terminations in t-PtBi$_2$. 
\textbf{a,b} Theoretical Fermi surfaces for KT and DH terminations. 
\textbf{c,d} Experimental Fermi surface maps ($h\nu$ = 17 eV, $T$ = 1.5 K) for the KT and DH terminations. 
\textbf{e,f} High-resolution laser Fermi surface maps ($h\nu$ = 6 eV, $T$ = 4 K) from the KT and DH terminations. 
\textbf{g,h} Zoomed-in Fermi arcs from KT and DH terminations in panels \textbf{a,b}.
\textbf{i,j} STM topography images ($T$ = 5 K) acquired from the KT and DH terminations. 
KT termination: $U$ = 0.01 V, $I$ = 200 pA; DH termination: $U$ = 0.05 V, $I$ = 400 pA. 
\textbf{k,l} DFT calculations of DH and KT terminations in $k_x$ and $k_y$ directions. 
\textbf{m,n} ARPES dispersion of the Fermi arcs for DH and KT terminations in $k_x$ and $k_y$ directions.}
\label{fig1}
\end{figure}

To unambiguously assign ARPES spectra to the correct termination, we performed a combination of STM topography, synchrotron and ultra-high resolution laser ARPES on the same samples. This allowed us to link the terminations to the electronic band structure. The results, together with theoretical calculations, are presented in Fig.~\ref{fig1}.

The Fermi surface maps for the two terminations, shown in Fig.~\ref{fig1}c,d, appear similar but exhibit subtle differences, most likely due to matrix element effects. In the DFT calculations (Figs.~\ref{fig1}a,b), the only notable difference between the two terminations lies in the shape and size of the Fermi arcs, while bulk remains unchanged.

Figures~\ref{fig1}(e,f) present laser Fermi surface maps of the Fermi arcs for both terminations. Thanks to the enhanced energy and momentum resolution provided by the 6~eV laser, compared to synchrotron measurements, we were able to clearly resolve the arc dispersions and directly compare them with theoretical predictions shown in Figs.~\ref{fig1}(g,h). Remarkably, the experimentally observed arcs closely match the calculated Fermi contours, capturing their shape in momentum space.

The Fermi arc dispersions along the $k_x$ and $k_y$ directions are shown in Figs.~\ref{fig1}(k-n), for both experiment and theory. Since Fermi arcs form an open contours, the dispersions along $k_x$ show simple electron-like behavior. In contrast, the $k_y$ dispersions cross the Fermi level only once and merge into the bulk continuum on the opposite side, resulting in an asymmetric profile.

The DH termination has good agreement with theoretical calculations. The arc bottom is located 6–8~meV below the Fermi level (see Figs.~\ref{fig1}k,m). The KT termination, however, displays significant deviations. First, the experimental Fermi velocity is substantially smaller compared to theory. Second, the dispersion exhibits an abrupt slope change near the arc bottom (see Fig.~\ref{fig1}n), accompanied by a suppression in photoemission intensity at that point. This kink-like feature points towards the presence of strong interactions, which we will explore in detail in Section~\ref{sec:flatband}.

\section{Van Hove singularity on DH termination}

\begin{figure}[H]%
\centering
\includegraphics[width=1.0\textwidth]{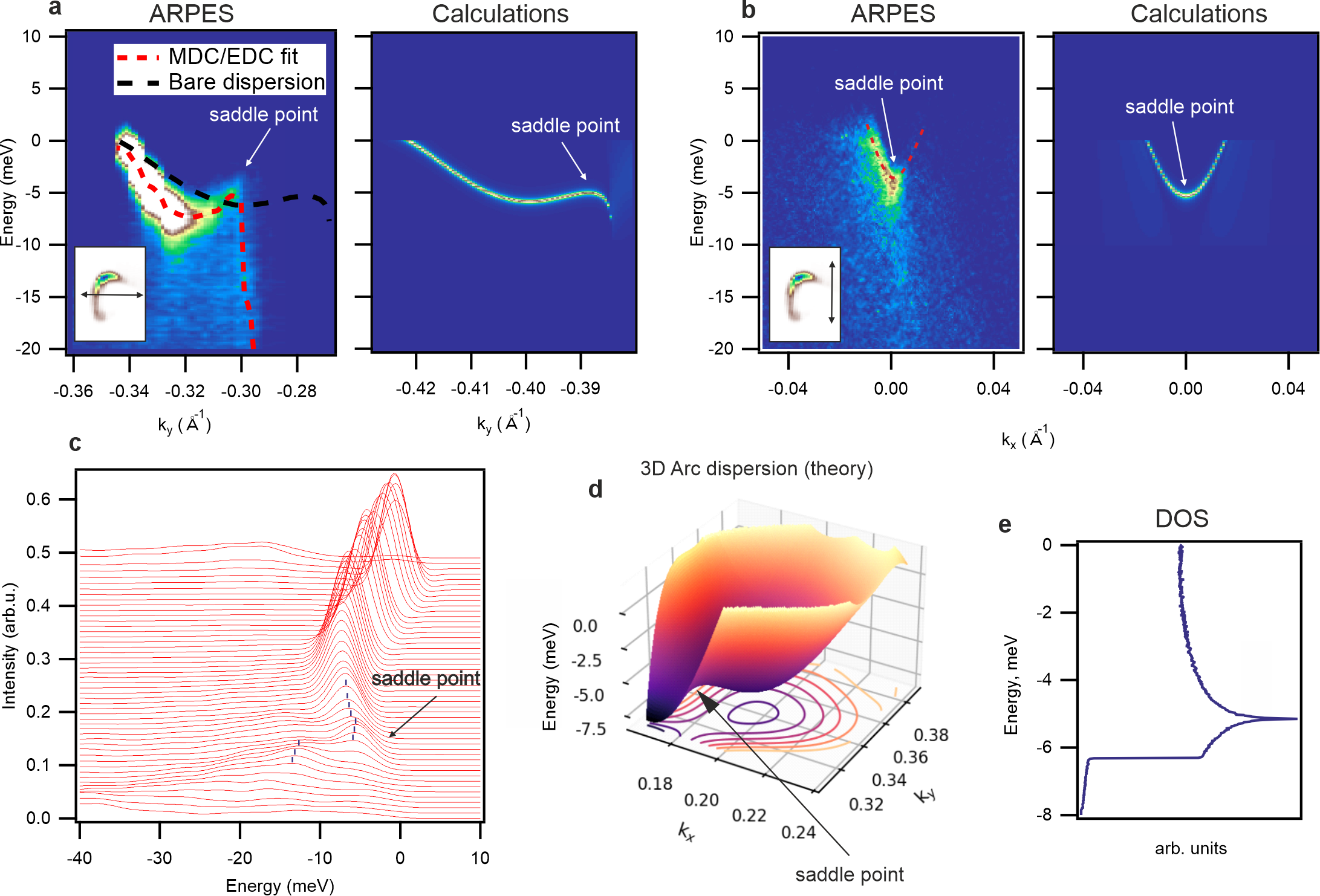}
\caption{Van Hove singularity in the Fermi arc on the DH termination. 
\textbf{a} ARPES and calculatedFermi arc dispersion along the line indicated on the Fermi surface map in the inset. 
\textbf{b} ARPES and calculated dispersion taken in the perpendicular direction from the point marked by the white arrow in \textbf{a}. The inset shows the direction of the cut on the Fermi surface map. 
\textbf{c} Smoothed EDCs extracted from \textbf{a}, showing the Van Hove saddle point. 
\textbf{d} Theoretical calculation of the 3D Fermi arc dispersion. 
\textbf{e} Density of states of the Fermi arc. A sharp peak at the energy of the saddle point indicates the presence of a VHS.}
\label{fig3}
\end{figure}

Detailed theoretical calculations reveal that the Fermi arc on DH terminations exhibits a saddle point in its band dispersion, as shown in  Fig.~\ref{fig3}d. Since the Fermi arc is a two-dimensional surface state, the presence of a saddle point is sufficient to induce a divergence in the DOS and create a VHS, as illustrated in DOS in Fig.~\ref{fig3}e.
A numerical analysis of the second derivatives at the saddle points confirms
that we encounter a type-I VHS, characterized by a logarithmic divergence in the
DOS.

Consistent with these theoretical predictions, our ARPES measurements also find evidence of the saddle point. Figure~\ref{fig3}a displays the Fermi arc dispersion, experimental and theoretical, along the $k_y$ direction, with red dashed lines indicating peak positions extracted from MDC fitting. In this direction, the dispersion near the white arrow exhibits a hole-like character. Figure~\ref{fig3}b shows the dispersion along the $k_y$ direction, perpendicular to the cut in Fig.~\ref{fig3}a, measured at the momentum position marked by the white arrow in the latter. Here, both in ARPES and calculations, the dispersion is electron-like, confirming the presence of a saddle point at the intersection of these two directions.

Further evidence of the saddle point is provided in Fig.~\ref{fig3}c, which shows smoothed EDCs corresponding to the hole-like dispersion in panel~a. The weak spectral intensity observed further below the Fermi level from the overlap of the Fermi arc with the bulk continuum, as also illustrated in the band structure calculations of the Fermi arc in right panel of Fig.~\ref{fig3}a.

The proximity of the VHS to the Fermi level is a highly favorable condition for high-$T_c$ superconductivity, as the associated divergence in the density of states can significantly strengthen pairing interactions. However, in the present measurements, the singularity is still too far below the Fermi level (about 5--7~meV), and its influence seemingly has not yet manifested.
This may be related to the fact that the superconducting gaps observed in our previous are limited to 1--2~meV.

In this context, it is worth nothing that a
logarithmic singularity leads to a BCS-like expression for
$T_c$, albeit with significant enhancement up to several orders of
magnitude~\cite{Hirish1986}. However, detailed numerical solutions of the Eliashberg
equations with realistic electron-phonon interactions reveal that such
enhancements occur only in close proximity to the VHS~\cite{Mahan1993}. When the chemical
potential lies more than about 10\% of the bandwidth away from the
singularity, the enhancement becomes negligible. A similar behavior is
predicted when superconductivity arises from purely repulsive
interactions, provided the singularity remains
logarithmic~\cite{Ojajrvi2024}.

 If the Fermi level could be tuned precisely to the energy of the singularity, a substantial enhancement in the gap size, and consequently in $T_c$ may be expected. In the following sections, we demonstrate that such Fermi level tuning is indeed achievable in this material.

\section{Flat band and strong renormalization on KT termination}\label{sec:flatband}
Strong coupling between Bloch electrons and another degree of freedom is considered one of the key ingredients in the formation of superconductivity. In many high-$T_c$ systems, such coupling leads to pronounced band renormalizations and kinks in the dispersion. In KT terminations of \ptbi, we find a particularly striking manifestation of such an effect.

\begin{figure}[H]%
\centering
\includegraphics[width=1.0\textwidth]{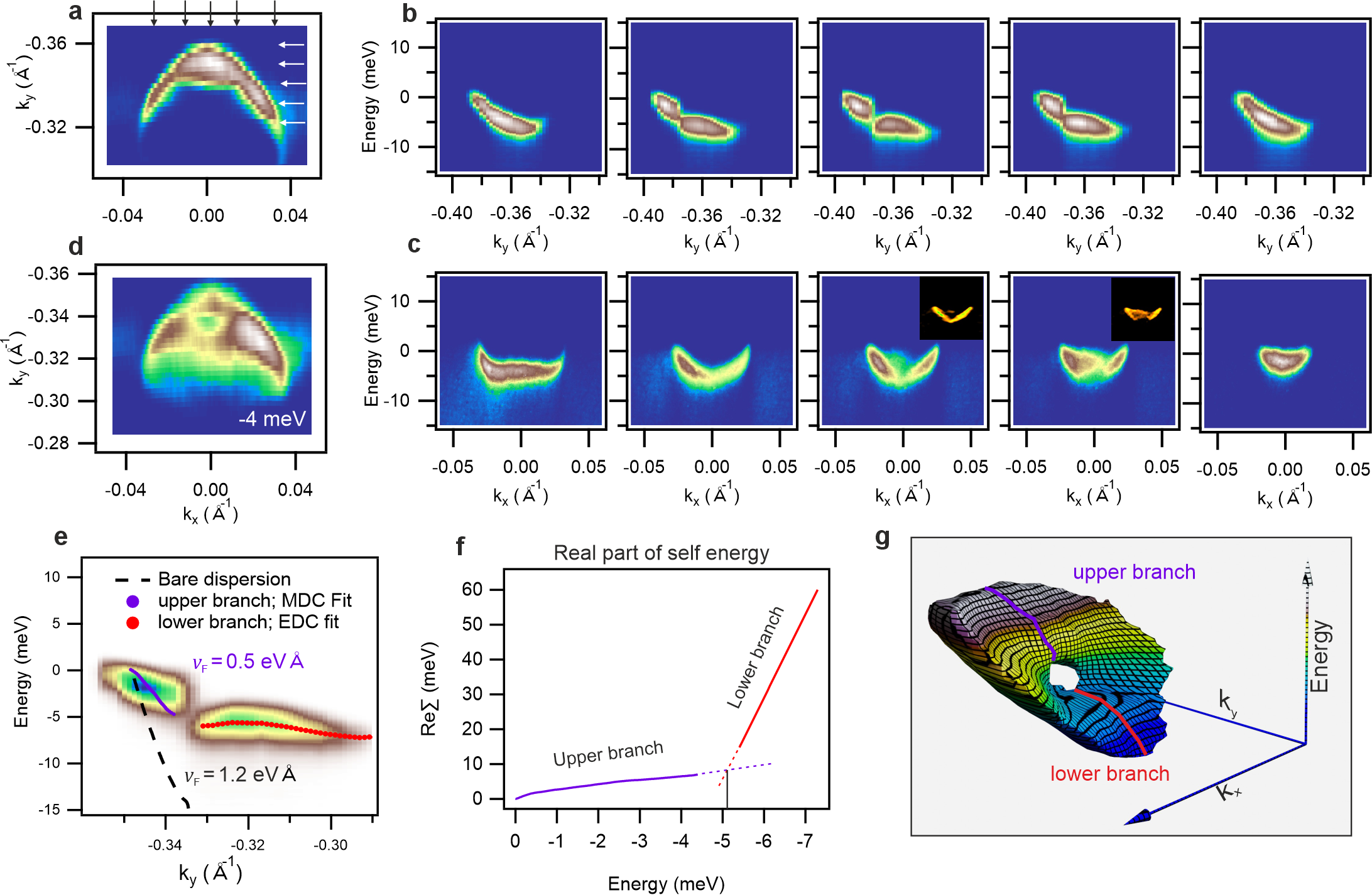}
\caption{Momentum-dependent strong coupling in the Fermi arc on the KT termination. 
\textbf{a} Fermi surface map of the Fermi arc. Black and white arrows indicate the directions of the ARPES dispersions shown in \textbf{b} and \textbf{c}, respectively. 
\textbf{b} Fermi arc dispersion measured along the black arrows in \textbf{a}. 
\textbf{c} Fermi arc dispersion measured along the white arrows in \textbf{a}. Insets show the second derivative of the corresponding images. 
\textbf{d} Constant-energy map taken $4 \pm 1$ meV below the Fermi level. 
\textbf{e} Central cut from \textbf{b} together with the MDC fit for the upper branch, EDC fit for the lower branch, and the DFT bare-band dispersion. 
\textbf{f} Real part of the self-energy.  
\textbf{g} 3D voxelgram of the Fermi arc, illustrating its dispersion in momentum and energy.}
\label{fig4}
\end{figure}

Figure~\ref{fig4}b presents a series of ARPES intensity distribution plots taken along the black arrows marked in Fermi map shown in Fig.~\ref{fig4}a. While the dispersions at the sides of the arc (far left and far right panels) appear conventional, the dispersions at the arc center reveal two distinct branches separated by a region of reduced spectral weight. This anomaly is further highlighted in the perpendicular cuts shown in Fig.~\ref{fig4}c, where the central panels exhibit a clear double-feature structure. Insets showing second-derivative images confirm the presence of two well-defined branches of  dispersion instead of a single continuous band.

We may attribute this splitting to a strong interaction between the surface electrons forming the arc and a bosonic mode, leading to a substantial renormalization of the electronic structure. In the simplest picture, such a coupling can produce a kink in the dispersion and a renormalization of the band. However, the effect in \ptbi\ is unusual in two respects: (i) the renormalization is strong enough to produce an almost dispersionless band segment, lower branch; and (ii) the coupling is strongly momentum dependent, being maximal at the arc center and rapidly vanishing towards its ends. 

Figure~\ref{fig4}e shows the central cut along $k_y$, together with the experimentally extracted dispersions and the corresponding bare-band dispersion from the calculations. To extract dispersion of the upper branch, we performed MDC fitting and determined the peak positions, while the lower, flat branch was traced using EDC peak positions due to the difficulty of MDC fitting in this region. The constant-energy map in Fig.~\ref{fig4}d, taken 4~meV below the Fermi level and 3D Voxel diagram of the arc Fermi surface in Fig.~\ref{fig4}g, further visualizes the spatial confinement of this interaction in momentum space. Notably, such a momentum-dependent renormalization  consistent with our recent observation of an anistropic superconducting gap on the Fermi arcs in this termination \cite{changdar_iwave}. This suggests that the same underlying interaction responsible for the band flattening here may also drive the anisotropic gap structure.

To quantify this coupling, we calculated the Fermi velocities for the bare Fermi arc, $v_F = 1.2$~eV\AA, and for the experimental dispersion of upper branch, $v_F = 0.5$~eV\AA, revealing a renormalization factor of approximately 2.5. In Fig.~\ref{fig4}f, we plot the real part of the self-energy for both the upper and lower branches. For the lower branch, the real part of the self-energy was obtained by fitting the EDC peak positions with a line and assuming a linear bare dispersion for the arc (see Supplementary Material).

Such a flat, interaction-renormalized band segment is a highly favorable condition for superconductivity as it boosts the DOS, similar to the effect of the VHS on DH termination.

% We also note that there is possibility that the KT termination host a van Hove singularity as well. However, we cannot provide definitive proof, since in theoretical calculations the arc dispersion merges with the bulk continuum. The experimental data may be interpreted as indicative of a van Hove singularity.For instance, the curvature of the lower branch in Fig.~\ref{fig4}e changes sign. Nevertheless, along the perpendicular momentum direction the dispersion remains too flat to make any reliable conclusions about the presence of a saddle point.

\section{Tuning capabilities}
In the previous sections, we demonstrated the presence of a van Hove singularity and a flat band on opposite terminations of \ptbi. On their own, these features are not sufficient to produce an enhanced superconductivity transition temperature: the increased density of states arising from the van Hove singularity or the flat band must lie sufficiently close to the Fermi level to strongly contribute to pairing. In the present case, both the van Hove singularity and the flat band are located approximately 5–8~meV below the Fermi level, whereas the superconducting gap measured by ARPES is only 1.5–2~meV. This suggests that these features have not yet played a dominant role in determining the $T_c$ and $\Delta$ observed in our ARPES experiments.

\begin{figure}[H]%
\centering
\includegraphics[width=1.0\textwidth]{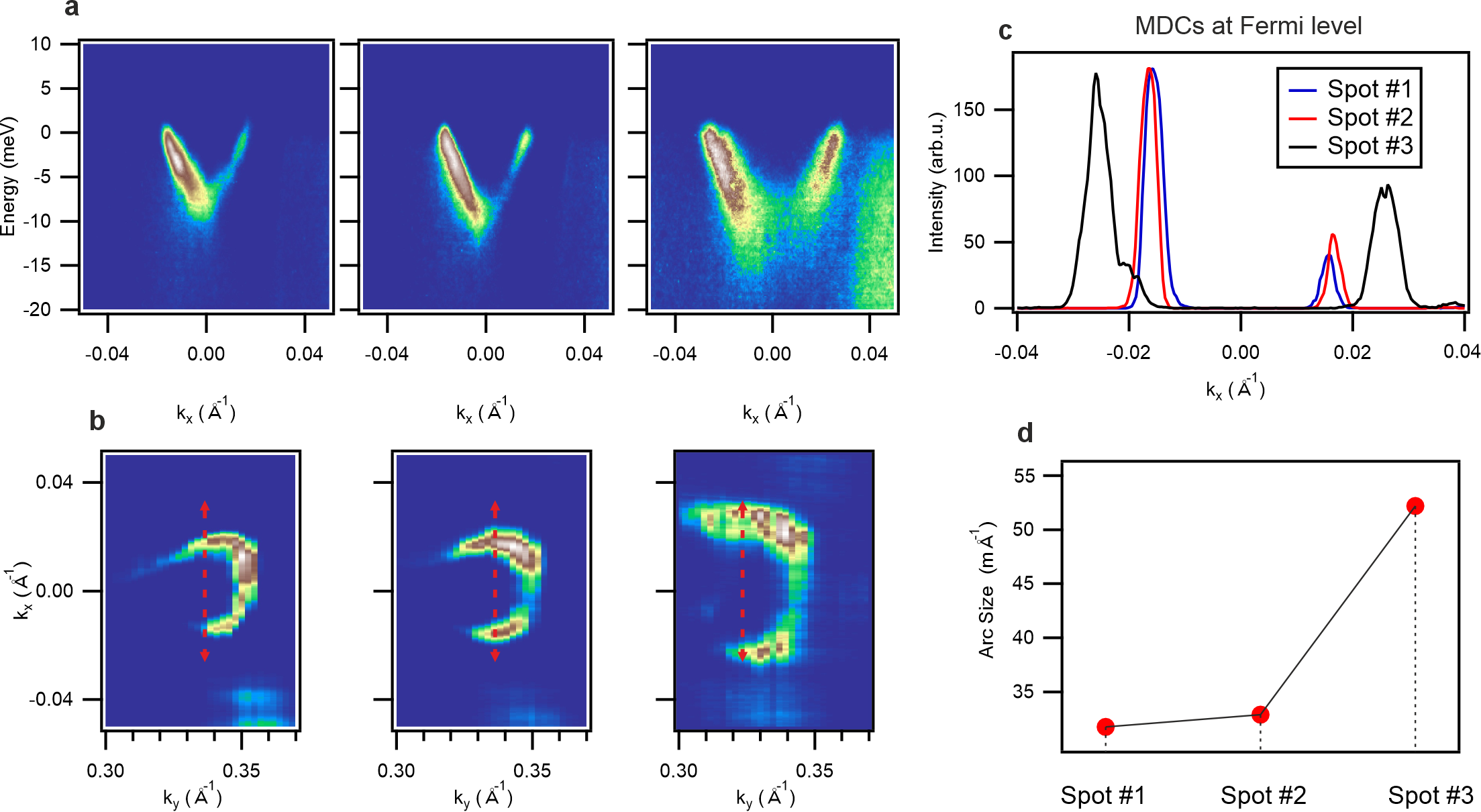}
\caption{
Change of the Fermi arc width with varying beam position on the sample surface. Spot \#3 was located close to the edge of the sample.
\textbf{a} ARPES dispersions of the Fermi arc from the DH termination at three different positions on the sample surface. 
\textbf{b} Fermi maps from the DH termination at three different positions on the sample surface. 
\textbf{c} Fermi-level MDCs corresponding to the images above.
\textbf{d} Size of the arc at three different spots}
\label{fig5}
\end{figure}

However, data shown in Fig.~\ref{fig5} suggests that the size of the Fermi arc is not homogeneous across the sample surface. By moving the light beam to different positions, we observed noticeable variations in the width of the Fermi arcs. Figures~\ref{fig5}a,b present ARPES momentum–intensity distributions together with Fermi surface maps measured at three distinct spots on the sample surface. Figures~\ref{fig5}c,d quantify the corresponding changes in arc width for these three locations. 

Remarkably, even with a beam spot size of approximately 40~$\mu$m, we could identify regions where the arc size differed significantly. This indicates the possible presence of local areas in which the van Hove singularity and the flat band are positioned much closer to the Fermi level, thereby creating a sufficiently high DOS to support the opening of superconducting gaps as large as 20~meV, as observed by STM \cite{Schimmel2024, XiaochunSTM}. Indeed, STM measurements show that such large gaps are not seen across the whole surface on every cleave \cite{Schimmel2024}. For instance, Fig.~3 in Ref.~\cite{Schimmel2024} reports superconducting gaps ranging from 0.5 to 3~meV at different locations, while Ref.~\cite{SpanishSTM} observes surface superconductivity in \ptbi\ with $T_c \approx 3$~K and $\Delta \approx 0.5$~meV. Owing to its atomically sharp tip, STM can selectively probe the most favorable regions on the surface, where the local Fermi level aligns the flat band or the van Hove singularity precisely with the Fermi level.

In summary, we have investigated the dispersion of the surface Fermi arcs in close proximity to the Fermi level. We find that the arc on the DH termination hosts a van Hove singularity, while the arc on the KT termination is strongly renormalized, pointing towards a coupling to a bosonic mode. If so, this coupling exhibits an unusual momentum dependence, being strongest at the center of the Fermi arc and diminishing towards its ends. Furthermore, we show that the size of the Fermi arc on the sample surface is spatially inhomogeneous, with sizable variations detected even when measuring ARPES using a 40~$\mu$m beam spot. Such variations could explain the spread in superconducting gap values reported by STM on \ptbi. Although the origin of this difference in arc size is currently unknown and has not yet been controlled, the ability to tune it in a reproducible manner would offer a powerful route to engineer the superconducting properties directly on the surface of a single stoichiometric material, without the need for complex heterostructures.

Taking into account a momentum-dependent superconducting gap consistent with $i$-wave pairing symmetry \cite{changdar_iwave}, the combination of a van Hove singularity, a momentum-selective flat band, and tunable size of the arc is particularly promising. Such a scenario could enhance unconventional surface superconductivity in \ptbi\ to regimes where it may host robust, topologically protected Majorana modes.

\section{Methods}
\subsection{Band structure calculations}

To investigate the electronic structure of PtBi$_2$, we performed
full-relativistic DFT calculations using the full-potential
local-orbital code FPLO~\cite{Koepernik1999} (version 22.01-63). For the
exchange-correlation term, the generalised gradient approximation (GGA)
was chosen~\cite{Perdew1997}. Convergence was achieved by sampling the BZ with a $12
\times 12 \times 12$ mesh of $k$-points. 
The experimental crystal structure from Ref.~\cite{ShipunovPRM} was used as structural
input. 

The surface Fermi arcs were obtained from the spectral densities of
semi-infinite slabs, computed via Green’s function
techniques~\cite{Kuibarov2024}. To resolve fine details the imaginary part of
the Green’s function is chosen as $10^{-4}$ eV.  For this purpose, we
constructed a Wannier function model using the dedicated FPLO
module~\cite{Koepernik2023} by projecting Kohn-Sham states onto a basis of 72
spin-orbitals that comprises Pt 6s, 5d, and Bi 6p states. The Wannier projection
is in good agreement with the DFT band structure from $-7$ to $5$~eV~\cite{Vocaturo2024}.

The 3D dispersion relation shown in Fig.~\ref{fig3}d and the associated DOS are obtained
extracting the points of maximal spectral intensity on a fine grid. Then to
construct a smooth function out of this data, the collection of points is then
approximated using the \texttt{SmoothBivariateSpline} routine from the Python package
\texttt{scipy.interpolate}.  

The 3D voxel diagram in Fig.~\ref{fig4}g was generated by selecting the most intense pixels (60--100\%) from the 3D Fermi map and rendering them as cubes. The terrain colormap is used to represent the energy scale.

\subsection{Experiment}
ARPES measurements were carried out at the $1^3$ ARPES end stations of the BESSY II synchrotron (Helmholtz-Zentrum Berlin) using 17~eV horizontally polarized light, as well as at the IFW Dresden laboratory with a continuous-wave 6~eV laser with horizontally polarized light. Samples were cleaved \textit{in situ} under a pressure lower than $1 \times 10^{-10}\,\mathrm{mbar}$ and measured at temperatures of 2~K at BESSY II synchrotron and 4~K at IFW Dresden laboratory. 

Characterization by STM was performed using an Omicron LT STM at $T=4.5$~K with electrochemically etched polycrystalline tungsten tips.

\bibliography{sn-bibliography}% common bib file


\begin{thebibliography}{32}
\ifx \bisbn   \undefined \def \bisbn  #1{ISBN #1}\fi
\ifx \binits  \undefined \def \binits#1{#1}\fi
\ifx \bauthor  \undefined \def \bauthor#1{#1}\fi
\ifx \batitle  \undefined \def \batitle#1{#1}\fi
\ifx \bjtitle  \undefined \def \bjtitle#1{#1}\fi
\ifx \bvolume  \undefined \def \bvolume#1{\textbf{#1}}\fi
\ifx \byear  \undefined \def \byear#1{#1}\fi
\ifx \bissue  \undefined \def \bissue#1{#1}\fi
\ifx \bfpage  \undefined \def \bfpage#1{#1}\fi
\ifx \blpage  \undefined \def \blpage #1{#1}\fi
\ifx \burl  \undefined \def \burl#1{\textsf{#1}}\fi
\ifx \doiurl  \undefined \def \doiurl#1{\url{https://doi.org/#1}}\fi
\ifx \betal  \undefined \def \betal{\textit{et al.}}\fi
\ifx \binstitute  \undefined \def \binstitute#1{#1}\fi
\ifx \binstitutionaled  \undefined \def \binstitutionaled#1{#1}\fi
\ifx \bctitle  \undefined \def \bctitle#1{#1}\fi
\ifx \beditor  \undefined \def \beditor#1{#1}\fi
\ifx \bpublisher  \undefined \def \bpublisher#1{#1}\fi
\ifx \bbtitle  \undefined \def \bbtitle#1{#1}\fi
\ifx \bedition  \undefined \def \bedition#1{#1}\fi
\ifx \bseriesno  \undefined \def \bseriesno#1{#1}\fi
\ifx \blocation  \undefined \def \blocation#1{#1}\fi
\ifx \bsertitle  \undefined \def \bsertitle#1{#1}\fi
\ifx \bsnm \undefined \def \bsnm#1{#1}\fi
\ifx \bsuffix \undefined \def \bsuffix#1{#1}\fi
\ifx \bparticle \undefined \def \bparticle#1{#1}\fi
\ifx \barticle \undefined \def \barticle#1{#1}\fi
\bibcommenthead
\ifx \bconfdate \undefined \def \bconfdate #1{#1}\fi
\ifx \botherref \undefined \def \botherref #1{#1}\fi
\ifx \url \undefined \def \url#1{\textsf{#1}}\fi
\ifx \bchapter \undefined \def \bchapter#1{#1}\fi
\ifx \bbook \undefined \def \bbook#1{#1}\fi
\ifx \bcomment \undefined \def \bcomment#1{#1}\fi
\ifx \oauthor \undefined \def \oauthor#1{#1}\fi
\ifx \citeauthoryear \undefined \def \citeauthoryear#1{#1}\fi
\ifx \endbibitem  \undefined \def \endbibitem {}\fi
\ifx \bconflocation  \undefined \def \bconflocation#1{#1}\fi
\ifx \arxivurl  \undefined \def \arxivurl#1{\textsf{#1}}\fi
\csname PreBibitemsHook\endcsname

%%% 1
\bibitem[\protect\citeauthoryear{Shimoyama and Motoki}{2023}]{Shimoyama2023}
\begin{barticle}
\bauthor{\bsnm{Shimoyama}, \binits{J.}},
\bauthor{\bsnm{Motoki}, \binits{T.}}:
\batitle{Current status of high temperature superconducting materials and their various applications}.
\bjtitle{IEEJ Transactions on Electrical and Electronic Engineering}
\bvolume{19}(\bissue{3}),
\bfpage{292}--\blpage{304}
(\byear{2023})
\doiurl{10.1002/tee.23976}
\end{barticle}
\endbibitem

%%% 2
\bibitem[\protect\citeauthoryear{Bok and Bouvier}{2012}]{Bok2012}
\begin{barticle}
\bauthor{\bsnm{Bok}, \binits{J.}},
\bauthor{\bsnm{Bouvier}, \binits{J.}}:
\batitle{Superconductivity and the van hove scenario}.
\bjtitle{Journal of Superconductivity and Novel Magnetism}
\bvolume{25}(\bissue{3}),
\bfpage{657}--\blpage{667}
(\byear{2012})
\doiurl{10.1007/s10948-012-1434-3}
\end{barticle}
\endbibitem

%%% 3
\bibitem[\protect\citeauthoryear{Borisenko}{2013}]{Borisenko2013NM}
\begin{barticle}
\bauthor{\bsnm{Borisenko}, \binits{S.}}:
\batitle{Fewer atoms, more information}.
\bjtitle{Nature Materials}
\bvolume{12}(\bissue{7}),
\bfpage{600}--\blpage{601}
(\byear{2013})
\doiurl{10.1038/nmat3683}
\end{barticle}
\endbibitem

%%% 4
\bibitem[\protect\citeauthoryear{Yokoya et~al.}{1996}]{Yokoya1996}
\begin{barticle}
\bauthor{\bsnm{Yokoya}, \binits{T.}},
\bauthor{\bsnm{Chainani}, \binits{A.}},
\bauthor{\bsnm{Takahashi}, \binits{T.}},
\bauthor{\bsnm{Katayama-Yoshida}, \binits{H.}},
\bauthor{\bsnm{Kasai}, \binits{M.}},
\bauthor{\bsnm{Tokura}, \binits{Y.}}:
\batitle{Extended van-hove singularity in sr2ruo4 observed by angle-resolved photoemission}.
\bjtitle{Physica C: Superconductivity}
\bvolume{263}(\bissue{1–4}),
\bfpage{505}--\blpage{509}
(\byear{1996})
\doiurl{10.1016/0921-4534(96)00073-1}
\end{barticle}
\endbibitem

%%% 5
\bibitem[\protect\citeauthoryear{Noad et~al.}{2023}]{Noad2023}
\begin{barticle}
\bauthor{\bsnm{Noad}, \binits{H.M.L.}},
\bauthor{\bsnm{Ishida}, \binits{K.}},
\bauthor{\bsnm{Li}, \binits{Y.-S.}},
\bauthor{\bsnm{Gati}, \binits{E.}},
\bauthor{\bsnm{Stangier}, \binits{V.}},
\bauthor{\bsnm{Kikugawa}, \binits{N.}},
\bauthor{\bsnm{Sokolov}, \binits{D.A.}},
\bauthor{\bsnm{Nicklas}, \binits{M.}},
\bauthor{\bsnm{Kim}, \binits{B.}},
\bauthor{\bsnm{Mazin}, \binits{I.I.}},
\bauthor{\bsnm{Garst}, \binits{M.}},
\bauthor{\bsnm{Schmalian}, \binits{J.}},
\bauthor{\bsnm{Mackenzie}, \binits{A.P.}},
\bauthor{\bsnm{Hicks}, \binits{C.W.}}:
\batitle{Giant lattice softening at a lifshitz transition in sr2ruo4}.
\bjtitle{Science}
\bvolume{382}(\bissue{6669}),
\bfpage{447}--\blpage{450}
(\byear{2023})
\doiurl{10.1126/science.adf3348}
\end{barticle}
\endbibitem

%%% 6
\bibitem[\protect\citeauthoryear{Kim et~al.}{2018}]{Kim2018}
\begin{barticle}
\bauthor{\bsnm{Kim}, \binits{K.}},
\bauthor{\bsnm{Kim}, \binits{S.}},
\bauthor{\bsnm{Kim}, \binits{J.S.}},
\bauthor{\bsnm{Kim}, \binits{H.}},
\bauthor{\bsnm{Park}, \binits{J.-H.}},
\bauthor{\bsnm{Min}, \binits{B.I.}}:
\batitle{Importance of the van hove singularity in superconducting ${\mathrm{pdte}}_{2}$}.
\bjtitle{Phys. Rev. B}
\bvolume{97},
\bfpage{165102}
(\byear{2018})
\doiurl{10.1103/PhysRevB.97.165102}
\end{barticle}
\endbibitem

%%% 7
\bibitem[\protect\citeauthoryear{Luo et~al.}{2023}]{Luo2023}
\begin{botherref}
\oauthor{\bsnm{Luo}, \binits{Y.}},
\oauthor{\bsnm{Han}, \binits{Y.}},
\oauthor{\bsnm{Liu}, \binits{J.}},
\oauthor{\bsnm{Chen}, \binits{H.}},
\oauthor{\bsnm{Huang}, \binits{Z.}},
\oauthor{\bsnm{Huai}, \binits{L.}},
\oauthor{\bsnm{Li}, \binits{H.}},
\oauthor{\bsnm{Wang}, \binits{B.}},
\oauthor{\bsnm{Shen}, \binits{J.}},
\oauthor{\bsnm{Ding}, \binits{S.}},
\oauthor{\bsnm{Li}, \binits{Z.}},
\oauthor{\bsnm{Peng}, \binits{S.}},
\oauthor{\bsnm{Wei}, \binits{Z.}},
\oauthor{\bsnm{Miao}, \binits{Y.}},
\oauthor{\bsnm{Sun}, \binits{X.}},
\oauthor{\bsnm{Ou}, \binits{Z.}},
\oauthor{\bsnm{Xiang}, \binits{Z.}},
\oauthor{\bsnm{Hashimoto}, \binits{M.}},
\oauthor{\bsnm{Lu}, \binits{D.}},
\oauthor{\bsnm{Yao}, \binits{Y.}},
\oauthor{\bsnm{Yang}, \binits{H.}},
\oauthor{\bsnm{Chen}, \binits{X.}},
\oauthor{\bsnm{Gao}, \binits{H.-J.}},
\oauthor{\bsnm{Qiao}, \binits{Z.}},
\oauthor{\bsnm{Wang}, \binits{Z.}},
\oauthor{\bsnm{He}, \binits{J.}}:
A unique van hove singularity in kagome superconductor csv3-xtaxsb5 with enhanced superconductivity.
Nature Communications
\textbf{14}(1)
(2023)
\doiurl{10.1038/s41467-023-39500-7}
\end{botherref}
\endbibitem

%%% 8
\bibitem[\protect\citeauthoryear{Villa-Cortés and De~la Peña-Seaman}{2022}]{VillaCorts2022}
\begin{barticle}
\bauthor{\bsnm{Villa-Cortés}, \binits{S.}},
\bauthor{\bsnm{Peña-Seaman}, \binits{O.}}:
\batitle{Effect of van hove singularity on the isotope effect and critical temperature of h3s hydride superconductor as a function of pressure}.
\bjtitle{Journal of Physics and Chemistry of Solids}
\bvolume{161},
\bfpage{110451}
(\byear{2022})
\doiurl{10.1016/j.jpcs.2021.110451}
\end{barticle}
\endbibitem

%%% 9
\bibitem[\protect\citeauthoryear{Li et~al.}{2009}]{Li2009}
\begin{barticle}
\bauthor{\bsnm{Li}, \binits{G.}},
\bauthor{\bsnm{Luican}, \binits{A.}},
\bauthor{\bsnm{Santos}, \binits{J.M.B.}},
\bauthor{\bsnm{Castro~Neto}, \binits{A.H.}},
\bauthor{\bsnm{Reina}, \binits{A.}},
\bauthor{\bsnm{Kong}, \binits{J.}},
\bauthor{\bsnm{Andrei}, \binits{E.Y.}}:
\batitle{Observation of van hove singularities in twisted graphene layers}.
\bjtitle{Nature Physics}
\bvolume{6}(\bissue{2}),
\bfpage{109}--\blpage{113}
(\byear{2009})
\doiurl{10.1038/nphys1463}
\end{barticle}
\endbibitem

%%% 10
\bibitem[\protect\citeauthoryear{Xu et~al.}{2021}]{Xu2021}
\begin{barticle}
\bauthor{\bsnm{Xu}, \binits{S.}},
\bauthor{\bsnm{Al~Ezzi}, \binits{M.M.}},
\bauthor{\bsnm{Balakrishnan}, \binits{N.}},
\bauthor{\bsnm{Garcia-Ruiz}, \binits{A.}},
\bauthor{\bsnm{Tsim}, \binits{B.}},
\bauthor{\bsnm{Mullan}, \binits{C.}},
\bauthor{\bsnm{Barrier}, \binits{J.}},
\bauthor{\bsnm{Xin}, \binits{N.}},
\bauthor{\bsnm{Piot}, \binits{B.A.}},
\bauthor{\bsnm{Taniguchi}, \binits{T.}},
\bauthor{\bsnm{Watanabe}, \binits{K.}},
\bauthor{\bsnm{Carvalho}, \binits{A.}},
\bauthor{\bsnm{Mishchenko}, \binits{A.}},
\bauthor{\bsnm{Geim}, \binits{A.K.}},
\bauthor{\bsnm{Fal’ko}, \binits{V.I.}},
\bauthor{\bsnm{Adam}, \binits{S.}},
\bauthor{\bsnm{Neto}, \binits{A.H.C.}},
\bauthor{\bsnm{Novoselov}, \binits{K.S.}},
\bauthor{\bsnm{Shi}, \binits{Y.}}:
\batitle{Tunable van hove singularities and correlated states in twisted monolayer–bilayer graphene}.
\bjtitle{Nature Physics}
\bvolume{17}(\bissue{5}),
\bfpage{619}--\blpage{626}
(\byear{2021})
\doiurl{10.1038/s41567-021-01172-9}
\end{barticle}
\endbibitem

%%% 11
\bibitem[\protect\citeauthoryear{Valla et~al.}{1999}]{Valla1999}
\begin{barticle}
\bauthor{\bsnm{Valla}, \binits{T.}},
\bauthor{\bsnm{Fedorov}, \binits{A.V.}},
\bauthor{\bsnm{Johnson}, \binits{P.D.}},
\bauthor{\bsnm{Wells}, \binits{B.O.}},
\bauthor{\bsnm{Hulbert}, \binits{S.L.}},
\bauthor{\bsnm{Li}, \binits{Q.}},
\bauthor{\bsnm{Gu}, \binits{G.D.}},
\bauthor{\bsnm{Koshizuka}, \binits{N.}}:
\batitle{Evidence for quantum critical behavior in the optimally doped cuprate bi2sr2cacu2o8+d}.
\bjtitle{Science}
\bvolume{285}(\bissue{5436}),
\bfpage{2110}--\blpage{2113}
(\byear{1999})
\doiurl{10.1126/science.285.5436.2110}
\end{barticle}
\endbibitem

%%% 12
\bibitem[\protect\citeauthoryear{Bostwick et~al.}{2010}]{Bostwick2010}
\begin{barticle}
\bauthor{\bsnm{Bostwick}, \binits{A.}},
\bauthor{\bsnm{Speck}, \binits{F.}},
\bauthor{\bsnm{Seyller}, \binits{T.}},
\bauthor{\bsnm{Horn}, \binits{K.}},
\bauthor{\bsnm{Polini}, \binits{M.}},
\bauthor{\bsnm{Asgari}, \binits{R.}},
\bauthor{\bsnm{MacDonald}, \binits{A.H.}},
\bauthor{\bsnm{Rotenberg}, \binits{E.}}:
\batitle{Observation of plasmarons in quasi-freestanding doped graphene}.
\bjtitle{Science}
\bvolume{328}(\bissue{5981}),
\bfpage{999}--\blpage{1002}
(\byear{2010})
\doiurl{10.1126/science.1186489}
\end{barticle}
\endbibitem

%%% 13
\bibitem[\protect\citeauthoryear{Lanzara et~al.}{2001}]{Lanzara2001}
\begin{barticle}
\bauthor{\bsnm{Lanzara}, \binits{A.}},
\bauthor{\bsnm{Bogdanov}, \binits{P.V.}},
\bauthor{\bsnm{Zhou}, \binits{X.J.}},
\bauthor{\bsnm{Kellar}, \binits{S.A.}},
\bauthor{\bsnm{Feng}, \binits{D.L.}},
\bauthor{\bsnm{Lu}, \binits{E.D.}},
\bauthor{\bsnm{Yoshida}, \binits{T.}},
\bauthor{\bsnm{Eisaki}, \binits{H.}},
\bauthor{\bsnm{Fujimori}, \binits{A.}},
\bauthor{\bsnm{Kishio}, \binits{K.}},
\bauthor{\bsnm{Shimoyama}, \binits{J.-I.}},
\bauthor{\bsnm{Noda}, \binits{T.}},
\bauthor{\bsnm{Uchida}, \binits{S.}},
\bauthor{\bsnm{Hussain}, \binits{Z.}},
\bauthor{\bsnm{Shen}, \binits{Z.-X.}}:
\batitle{Evidence for ubiquitous strong electron–phonon coupling in high-temperature superconductors}.
\bjtitle{Nature}
\bvolume{412}(\bissue{6846}),
\bfpage{510}--\blpage{514}
(\byear{2001})
\doiurl{10.1038/35087518}
\end{barticle}
\endbibitem

%%% 14
\bibitem[\protect\citeauthoryear{Dahm et~al.}{2009}]{Dahm2009}
\begin{barticle}
\bauthor{\bsnm{Dahm}, \binits{T.}},
\bauthor{\bsnm{Hinkov}, \binits{V.}},
\bauthor{\bsnm{Borisenko}, \binits{S.V.}},
\bauthor{\bsnm{Kordyuk}, \binits{A.A.}},
\bauthor{\bsnm{Zabolotnyy}, \binits{V.B.}},
\bauthor{\bsnm{Fink}, \binits{J.}},
\bauthor{\bsnm{B\"{u}chner}, \binits{B.}},
\bauthor{\bsnm{Scalapino}, \binits{D.J.}},
\bauthor{\bsnm{Hanke}, \binits{W.}},
\bauthor{\bsnm{Keimer}, \binits{B.}}:
\batitle{Strength of the spin-fluctuation-mediated pairing interaction in a high-temperature superconductor}.
\bjtitle{Nature Physics}
\bvolume{5}(\bissue{3}),
\bfpage{217}--\blpage{221}
(\byear{2009})
\doiurl{10.1038/nphys1180}
\end{barticle}
\endbibitem

%%% 15
\bibitem[\protect\citeauthoryear{Kordyuk et~al.}{2005}]{Kordyuk2005}
\begin{barticle}
\bauthor{\bsnm{Kordyuk}, \binits{A.A.}},
\bauthor{\bsnm{Borisenko}, \binits{S.V.}},
\bauthor{\bsnm{Koitzsch}, \binits{A.}},
\bauthor{\bsnm{Fink}, \binits{J.}},
\bauthor{\bsnm{Knupfer}, \binits{M.}},
\bauthor{\bsnm{Berger}, \binits{H.}}:
\batitle{Bare electron dispersion from experiment: Self-consistent self-energy analysis of photoemission data}.
\bjtitle{Phys. Rev. B}
\bvolume{71},
\bfpage{214513}
(\byear{2005})
\doiurl{10.1103/PhysRevB.71.214513}
\end{barticle}
\endbibitem

%%% 16
\bibitem[\protect\citeauthoryear{Kuibarov et~al.}{2024}]{Kuibarov2024}
\begin{barticle}
\bauthor{\bsnm{Kuibarov}, \binits{A.}},
\bauthor{\bsnm{Suvorov}, \binits{O.}},
\bauthor{\bsnm{Vocaturo}, \binits{R.}},
\bauthor{\bsnm{Fedorov}, \binits{A.}},
\bauthor{\bsnm{Lou}, \binits{R.}},
\bauthor{\bsnm{Merkwitz}, \binits{L.}},
\bauthor{\bsnm{Voroshnin}, \binits{V.}},
\bauthor{\bsnm{Facio}, \binits{J.I.}},
\bauthor{\bsnm{Koepernik}, \binits{K.}},
\bauthor{\bsnm{Yaresko}, \binits{A.}},
\bauthor{\bsnm{Shipunov}, \binits{G.}},
\bauthor{\bsnm{Aswartham}, \binits{S.}},
\bauthor{\bsnm{{van den Brink}}, \binits{J.}},
\bauthor{\bsnm{B\"{u}chner}, \binits{B.}},
\bauthor{\bsnm{Borisenko}, \binits{S.}}:
\batitle{Evidence of superconducting fermi arcs}.
\bjtitle{Nature}
\bvolume{626}(\bissue{7998}),
\bfpage{294}--\blpage{299}
(\byear{2024})
\doiurl{10.1038/s41586-023-06977-7}
\end{barticle}
\endbibitem

%%% 17
\bibitem[\protect\citeauthoryear{Schimmel et~al.}{2024}]{Schimmel2024}
\begin{botherref}
\oauthor{\bsnm{Schimmel}, \binits{S.}},
\oauthor{\bsnm{Fasano}, \binits{Y.}},
\oauthor{\bsnm{Hoffmann}, \binits{S.}},
\oauthor{\bsnm{Besproswanny}, \binits{J.}},
\oauthor{\bsnm{Corredor~Bohorquez}, \binits{L.T.}},
\oauthor{\bsnm{Puig}, \binits{J.}},
\oauthor{\bsnm{Elshalem}, \binits{B.-C.}},
\oauthor{\bsnm{Kalisky}, \binits{B.}},
\oauthor{\bsnm{Shipunov}, \binits{G.}},
\oauthor{\bsnm{Baumann}, \binits{D.}},
\oauthor{\bsnm{Aswartham}, \binits{S.}},
\oauthor{\bsnm{B\"{u}chner}, \binits{B.}},
\oauthor{\bsnm{Hess}, \binits{C.}}:
Surface superconductivity in the topological weyl semimetal t-ptbi2.
Nature Communications
\textbf{15}(1)
(2024)
\doiurl{10.1038/s41467-024-54389-6}
\end{botherref}
\endbibitem

%%% 18
\bibitem[\protect\citeauthoryear{Changdar et~al.}{2025}]{changdar_iwave}
\begin{botherref}
\oauthor{\bsnm{Changdar}, \binits{S.}},
\oauthor{\bsnm{Suvorov}, \binits{O.}},
\oauthor{\bsnm{Kuibarov}, \binits{A.}},
\oauthor{\bsnm{Thirupathaiah}, \binits{S.}},
\oauthor{\bsnm{Shipunov}, \binits{G.}},
\oauthor{\bsnm{Aswartham}, \binits{S.}},
\oauthor{\bsnm{Wurmehl}, \binits{S.}},
\oauthor{\bsnm{Kovalchuk}, \binits{I.}},
\oauthor{\bsnm{Koepernik}, \binits{K.}},
\oauthor{\bsnm{Timm}, \binits{C.}},
\oauthor{\bsnm{B\"{u}chner}, \binits{B.}},
\oauthor{\bsnm{Fulga}, \binits{I.C.}},
\oauthor{\bsnm{Borisenko}, \binits{S.}},
\oauthor{\bsnm{{van den Brink}}, \binits{J.}}:
Topological nodal $i$-wave superconductivity in PtBi$_2$.
arXiv
(2025).
\doiurl{10.48550/ARXIV.2507.01774} .
\url{https://arxiv.org/abs/2507.01774}
\end{botherref}
\endbibitem

%%% 19
\bibitem[\protect\citeauthoryear{Besproswanny et~al.}{2025}]{JuliaTemperatureSTM}
\begin{botherref}
\oauthor{\bsnm{Besproswanny}, \binits{J.}},
\oauthor{\bsnm{Schimmel}, \binits{S.}},
\oauthor{\bsnm{Fasano}, \binits{Y.}},
\oauthor{\bsnm{Shipunov}, \binits{G.}},
\oauthor{\bsnm{Aswartham}, \binits{S.}},
\oauthor{\bsnm{Baumann}, \binits{D.}},
\oauthor{\bsnm{B\"{u}chner}, \binits{B.}},
\oauthor{\bsnm{Hess}, \binits{C.}}:
Temperature dependence of surface superconductivity in t-PtBi$_2$.
arXiv
(2025).
\doiurl{10.48550/ARXIV.2507.10187} .
\url{https://arxiv.org/abs/2507.10187}
\end{botherref}
\endbibitem

%%% 20
\bibitem[\protect\citeauthoryear{Huang et~al.}{2025}]{XiaochunSTM}
\begin{botherref}
\oauthor{\bsnm{Huang}, \binits{X.}},
\oauthor{\bsnm{Zhao}, \binits{L.}},
\oauthor{\bsnm{Schimmel}, \binits{S.}},
\oauthor{\bsnm{Besproswanny}, \binits{J.}},
\oauthor{\bsnm{H\"{a}rtl}, \binits{P.}},
\oauthor{\bsnm{Hess}, \binits{C.}},
\oauthor{\bsnm{B\"{u}chner}, \binits{B.}},
\oauthor{\bsnm{Bode}, \binits{M.}}:
Sizable superconducting gap and anisotropic chiral topological superconductivity in the Weyl semimetal PtBi$_2$.
arXiv
(2025).
\doiurl{10.48550/ARXIV.2507.13843} .
\url{https://arxiv.org/abs/2507.13843}
\end{botherref}
\endbibitem

%%% 21
\bibitem[\protect\citeauthoryear{Hoffmann et~al.}{2024}]{Hoffmann2024}
\begin{botherref}
\oauthor{\bsnm{Hoffmann}, \binits{S.}},
\oauthor{\bsnm{Schimmel}, \binits{S.}},
\oauthor{\bsnm{Vocaturo}, \binits{R.}},
\oauthor{\bsnm{Puig}, \binits{J.}},
\oauthor{\bsnm{Shipunov}, \binits{G.}},
\oauthor{\bsnm{Janson}, \binits{O.}},
\oauthor{\bsnm{Aswartham}, \binits{S.}},
\oauthor{\bsnm{Baumann}, \binits{D.}},
\oauthor{\bsnm{B\"{u}chner}, \binits{B.}},
\oauthor{\bsnm{{van den Brink}}, \binits{J.}},
\oauthor{\bsnm{Fasano}, \binits{Y.}},
\oauthor{\bsnm{Facio}, \binits{J.I.}},
\oauthor{\bsnm{Hess}, \binits{C.}}:
Fermi arcs dominating the electronic surface properties of trigonal ptbi2.
Advanced Physics Research
\textbf{4}(5)
(2024)
\doiurl{10.1002/apxr.202400150}
\end{botherref}
\endbibitem

%%% 22
\bibitem[\protect\citeauthoryear{Veyrat et~al.}{2023}]{Veyrat2023}
\begin{barticle}
\bauthor{\bsnm{Veyrat}, \binits{A.}},
\bauthor{\bsnm{Labracherie}, \binits{V.}},
\bauthor{\bsnm{Bashlakov}, \binits{D.L.}},
\bauthor{\bsnm{Caglieris}, \binits{F.}},
\bauthor{\bsnm{Facio}, \binits{J.I.}},
\bauthor{\bsnm{Shipunov}, \binits{G.}},
\bauthor{\bsnm{Charvin}, \binits{T.}},
\bauthor{\bsnm{Acharya}, \binits{R.}},
\bauthor{\bsnm{Naidyuk}, \binits{Y.}},
\bauthor{\bsnm{Giraud}, \binits{R.}},
\bauthor{\bsnm{{van den Brink}}, \binits{J.}},
\bauthor{\bsnm{B\"{u}chner}, \binits{B.}},
\bauthor{\bsnm{Hess}, \binits{C.}},
\bauthor{\bsnm{Aswartham}, \binits{S.}},
\bauthor{\bsnm{Dufouleur}, \binits{J.}}:
\batitle{Berezinskii–kosterlitz–thouless transition in the type-i weyl semimetal ptbi2}.
\bjtitle{Nano Letters}
\bvolume{23}(\bissue{4}),
\bfpage{1229}--\blpage{1235}
(\byear{2023})
\doiurl{10.1021/acs.nanolett.2c04297}
\end{barticle}
\endbibitem

%%% 23
\bibitem[\protect\citeauthoryear{Shipunov et~al.}{2020}]{ShipunovPRM}
\begin{barticle}
\bauthor{\bsnm{Shipunov}, \binits{G.}},
\bauthor{\bsnm{Kovalchuk}, \binits{I.}},
\bauthor{\bsnm{Piening}, \binits{B.R.}},
\bauthor{\bsnm{Labracherie}, \binits{V.}},
\bauthor{\bsnm{Veyrat}, \binits{A.}},
\bauthor{\bsnm{Wolf}, \binits{D.}},
\bauthor{\bsnm{Lubk}, \binits{A.}},
\bauthor{\bsnm{Subakti}, \binits{S.}},
\bauthor{\bsnm{Giraud}, \binits{R.}},
\bauthor{\bsnm{Dufouleur}, \binits{J.}},
\bauthor{\bsnm{Shokri}, \binits{S.}},
\bauthor{\bsnm{Caglieris}, \binits{F.}},
\bauthor{\bsnm{Hess}, \binits{C.}},
\bauthor{\bsnm{Efremov}, \binits{D.V.}},
\bauthor{\bsnm{B\"uchner}, \binits{B.}},
\bauthor{\bsnm{Aswartham}, \binits{S.}}:
\batitle{Polymorphic ${\mathrm{ptbi}}_{2}$: Growth, structure, and superconducting properties}.
\bjtitle{Phys. Rev. Mater.}
\bvolume{4},
\bfpage{124202}
(\byear{2020})
\doiurl{10.1103/PhysRevMaterials.4.124202}
\end{barticle}
\endbibitem

%%% 24
\bibitem[\protect\citeauthoryear{Majchrzak et~al.}{2025}]{PtBi2_pumpProbe}
\begin{barticle}
\bauthor{\bsnm{Majchrzak}, \binits{P.}},
\bauthor{\bsnm{Sanders}, \binits{C.}},
\bauthor{\bsnm{Zhang}, \binits{Y.}},
\bauthor{\bsnm{Kuibarov}, \binits{A.}},
\bauthor{\bsnm{Suvorov}, \binits{O.}},
\bauthor{\bsnm{Springate}, \binits{E.}},
\bauthor{\bsnm{Kovalchuk}, \binits{I.}},
\bauthor{\bsnm{Aswartham}, \binits{S.}},
\bauthor{\bsnm{Shipunov}, \binits{G.}},
\bauthor{\bsnm{B\"uchner}, \binits{B.}},
\bauthor{\bsnm{Yaresko}, \binits{A.}},
\bauthor{\bsnm{Borisenko}, \binits{S.}},
\bauthor{\bsnm{Hofmann}, \binits{P.}}:
\batitle{Machine-learning approach to understanding ultrafast carrier dynamics in the three-dimensional brillouin zone of ${\mathrm{ptbi}}_{2}$}.
\bjtitle{Phys. Rev. Res.}
\bvolume{7},
\bfpage{013025}
(\byear{2025})
\doiurl{10.1103/PhysRevResearch.7.013025}
\end{barticle}
\endbibitem

%%% 25
\bibitem[\protect\citeauthoryear{Vocaturo et~al.}{2024}]{Vocaturo2024}
\begin{barticle}
\bauthor{\bsnm{Vocaturo}, \binits{R.}},
\bauthor{\bsnm{Koepernik}, \binits{K.}},
\bauthor{\bsnm{Facio}, \binits{J.I.}},
\bauthor{\bsnm{Timm}, \binits{C.}},
\bauthor{\bsnm{Fulga}, \binits{I.C.}},
\bauthor{\bsnm{Janson}, \binits{O.}},
\bauthor{\bsnm{Brink}, \binits{J.}}:
\batitle{Electronic structure of the surface-superconducting weyl semimetal ${\mathrm{ptbi}}_{2}$}.
\bjtitle{Phys. Rev. B}
\bvolume{110},
\bfpage{054504}
(\byear{2024})
\doiurl{10.1103/PhysRevB.110.054504}
\end{barticle}
\endbibitem

%%% 26
\bibitem[\protect\citeauthoryear{Hirsch and Scalapino}{1986}]{Hirish1986}
\begin{barticle}
\bauthor{\bsnm{Hirsch}, \binits{J.E.}},
\bauthor{\bsnm{Scalapino}, \binits{D.J.}}:
\batitle{Enhanced superconductivity in quasi two-dimensional systems}.
\bjtitle{Phys. Rev. Lett.}
\bvolume{56},
\bfpage{2732}--\blpage{2735}
(\byear{1986})
\doiurl{10.1103/PhysRevLett.56.2732}
\end{barticle}
\endbibitem

%%% 27
\bibitem[\protect\citeauthoryear{Mahan}{1993}]{Mahan1993}
\begin{barticle}
\bauthor{\bsnm{Mahan}, \binits{G.D.}}:
\batitle{Electron-phonon interaction near van hove singularities}.
\bjtitle{Phys. Rev. B}
\bvolume{48},
\bfpage{16557}--\blpage{16563}
(\byear{1993})
\doiurl{10.1103/PhysRevB.48.16557}
\end{barticle}
\endbibitem

%%% 28
\bibitem[\protect\citeauthoryear{Ojaj\"{a}rvi et~al.}{2024}]{Ojajrvi2024}
\begin{botherref}
\oauthor{\bsnm{Ojaj\"{a}rvi}, \binits{R.}},
\oauthor{\bsnm{Chubukov}, \binits{A.V.}},
\oauthor{\bsnm{Lee}, \binits{Y.-C.}},
\oauthor{\bsnm{Garst}, \binits{M.}},
\oauthor{\bsnm{Schmalian}, \binits{J.}}:
Pairing at a single van hove point.
npj Quantum Materials
\textbf{9}(1)
(2024)
\doiurl{10.1038/s41535-024-00717-4}
\end{botherref}
\endbibitem

%%% 29
\bibitem[\protect\citeauthoryear{Moreno et~al.}{2025}]{SpanishSTM}
\begin{botherref}
\oauthor{\bsnm{Moreno}, \binits{J.A.}},
\oauthor{\bsnm{Talavera}, \binits{P.G.}},
\oauthor{\bsnm{Herrera}, \binits{E.}},
\oauthor{\bsnm{Valle}, \binits{S.L.}},
\oauthor{\bsnm{Li}, \binits{Z.}},
\oauthor{\bsnm{Wang}, \binits{L.-L.}},
\oauthor{\bsnm{Bud'ko}, \binits{S.}},
\oauthor{\bsnm{Buzdin}, \binits{A.I.}},
\oauthor{\bsnm{Guillamón}, \binits{I.}},
\oauthor{\bsnm{Canfield}, \binits{P.C.}},
\oauthor{\bsnm{Suderow}, \binits{H.}}:
Robust surface superconductivity and vortex lattice in the Weyl semimetal t-PtBi$_2$.
arXiv
(2025).
\doiurl{10.48550/ARXIV.2508.04867} .
\url{https://arxiv.org/abs/2508.04867}
\end{botherref}
\endbibitem

%%% 30
\bibitem[\protect\citeauthoryear{Koepernik and Eschrig}{1999}]{Koepernik1999}
\begin{barticle}
\bauthor{\bsnm{Koepernik}, \binits{K.}},
\bauthor{\bsnm{Eschrig}, \binits{H.}}:
\batitle{Full-potential nonorthogonal local-orbital minimum-basis band-structure scheme}.
\bjtitle{Phys. Rev. B}
\bvolume{59}(\bissue{3}),
\bfpage{1743}--\blpage{1757}
(\byear{1999})
\doiurl{10.1103/physrevb.59.1743}
\end{barticle}
\endbibitem

%%% 31
\bibitem[\protect\citeauthoryear{Perdew et~al.}{1997}]{Perdew1997}
\begin{barticle}
\bauthor{\bsnm{Perdew}, \binits{J.P.}},
\bauthor{\bsnm{Burke}, \binits{K.}},
\bauthor{\bsnm{Ernzerhof}, \binits{M.}}:
\batitle{Generalized gradient approximation made simple}.
\bjtitle{Phys. Rev. Lett.}
\bvolume{78},
\bfpage{1396}--\blpage{1396}
(\byear{1997})
\doiurl{10.1103/PhysRevLett.78.1396}
\end{barticle}
\endbibitem

%%% 32
\bibitem[\protect\citeauthoryear{Koepernik et~al.}{2023}]{Koepernik2023}
\begin{barticle}
\bauthor{\bsnm{Koepernik}, \binits{K.}},
\bauthor{\bsnm{Janson}, \binits{O.}},
\bauthor{\bsnm{Sun}, \binits{Y.}},
\bauthor{\bsnm{Brink}, \binits{J.}}:
\batitle{{Symmetry-conserving maximally projected {Wannier} functions}}.
\bjtitle{Phys. Rev. B}
\bvolume{107},
\bfpage{235135}
(\byear{2023})
\doiurl{10.1103/PhysRevB.107.235135}
\end{barticle}
\endbibitem

\end{thebibliography}
\clearpage

% \begin{figure}[H]%
% \centering
% \includegraphics[width=1.0\textwidth]{Figures/Fig2.png}
% \caption{Fermi arc dispersion from two terminations. \textbf{a,b,c,d} Theoretical calculations of the Fermi arc dispersions. The direction of dispersion in momentum space corresponds to that in the respective experimental images below. \textbf{e,f,g,h} Experimental Fermi arc dispersions from the DH and Kagome terminations, measured along the directions indicated on the Fermi surface maps in the insets.}
% \label{fig2}
% \end{figure}

\end{document}